# NUMERICAL SOLUTION OF MANY-BODY WAVE SCATTERING PROBLEM FOR SMALL PARTICLES

M. I. Andriychuk[1], A. G. Ramm[2]

[1]Institute of Applied Problems of Mechanics and Mathematics, NASU
3"B" Naukova St., 79060, Lviv, Ukraine, e-mail: andr@iapmm.lviv.ua
[2]Mathematics Department, Kansas State University, Manhattan, KS, 66506, USA
e-mail: ramm@math.ksu.edu

*Abstract.* - A numerical approach to the problem of wave scattering by many small particles is developed under the assumptions $ka \ll 1$, $d \gg a$, where $a$ is the size of the particles and $d$ is the distance between the neighboring particles. On the wavelength one may have many small particles. An impedance boundary conditions are assumed on the boundaries of small particles. The results of numerical simulation show good agreement with the theory. They open a way to numerical simulation of the method for creating materials with a desired refraction coefficient.

## 1. Introduction

Theory of wave scattering by small particles of arbitrary shapes was developed by A. G. Ramm in [1] (see also [2]), where analytical formulas for $S$-matrix for acoustic and electromagnetic wave scattering by small bodies are derived. These formulas allow one to calculate the $S$-matrix with arbitrary accuracy and can be used in many practical problems. An asymptotically exact solution of the many-body wave scattering problem was developed in [3] under the assumptions $ka \ll 1$, $d = O(a^{1/3})$, $M = O(1/a)$, where $a$ is the size of the particles, $k = 2\pi/\lambda$ is the wave number, $d$ is the distance between neighboring particles, and $M$ is the total number of the particles embedded in a bounded domain $D \subset R^3$. An impedance boundary condition on the boundary $S_m$ of the $m$-th particle $D_m$ was assumed. In [4] the assumptions are more general:

$$\zeta_m = \frac{h(x_m)}{a^\kappa}, \ d = O(a^{(2-\kappa)/3}), \ M = O(\frac{1}{a^{2-\kappa}}), \ \kappa \in (0,1) \qquad (1)$$

where $\zeta_m$ is the boundary impedance, $h_m = h(x_m)$, $x_m \in D_m$, and $h(x) \in C(D)$ is an arbitrary continuous in $\overline{D}$ function, $\operatorname{Im} h \leq 0$.

The initial field $u_0$ satisfies the Helmholtz equation in $R^3$ and the scattered field satisfies the radiation condition. We assume in this paper that $\kappa \in (0,1)$ and the small particle $D_m$ is a ball of radius $a$ centered at the point $x_m$, $1 \leq m \leq M$.

## 2. The solution of scattering problem

The scattering problem is

$$[\nabla^2 + k^2 n_0^2(x)]u_M = 0 \ \text{in} \ R^3 \setminus \bigcup_{m=1}^M D_m, \qquad (2)$$

$$\frac{\partial u_M}{\partial N} = \zeta_m u_M \text{ on } S_m, \ 1 \leq m \leq M, \qquad (3)$$

where

$$u_M = u_0 + v_M, \qquad (4)$$

$u_0$ is solution to (2), (3), (4) with $M = 0$, (i.e. in the absence of embedded particles) and with the incident field $e^{ik\alpha \cdot x}$, and $v_M$ satisfies the radiation conditions.

It was proved in [3] that unique solution to (2) - (4) is of the form

$$u_M(x) = u_0(x) + \sum_{m=1}^{M} \int_{S_m} G(x,y)\sigma_m(y)dy, \qquad (5)$$

where $G(x,y)$ is Green's function of Helmholtz equation in the case when $M = 0$.

Let us define the "effective field" $u_e$, acting on the $m$-th particle:

$$u_e(x) := u_e(x,a) := u_e^{(m)}(x) := u_M(x) - \int_{S_m} G(x,y)\sigma_m(y)dy, \ x \in R^3, \qquad (6)$$

The function $\sigma_m(y)$ solves an exact integral equation, which is solved asymptotically in [4] as $a \to 0$. Let $h(x) \in C(D)$, $\text{Im}\, h \leq 0$, be arbitrary. Let $\Delta_p \subset D$ be any subdomain of $D$, and $N(\Delta_p)$ be the number of particles in $\Delta_p$. We assume that

$$N(\Delta_p) = \frac{1}{a^{2-\kappa}} \int_{\Delta_p} N(x)dx[1 + o(1)], \ a \to 0, \qquad (7)$$

where $N(x) \geq 0$ is a given continuous function in $D$. The following result was proved in [4] (Theorem 1): there exists the limit $u(x)$ of $u_e(x)$ as $a \to 0$:

$$\lim_{a \to 0} \|u_e(x) - u(x)\|_{C(D)} = 0, \qquad (8)$$

and $u(x)$ solves the following equation:

$$u(x) = u_0(x) - 4\pi \int_D G(x,y)h(y)N(y)u(y)dy. \qquad (9)$$

This is the equation, derived in [4] for the limiting effective field in the medium, created by embedding many small particles with the distribution law (7).

## 3. Approximate representation of effective field

Let us derive an explicit formula for the effective field $u_e$. Rewrite the exact formula (5) as:

$$u_M(x) = u_0(x) + \sum_{m=1}^{M} G(x,x_m)Q_m + \sum_{m=1}^{M} \int_{S_m} [G(x,y) - G(x,x_m)]\sigma_m(y)dy, \qquad (10)$$

where

$$Q_m = \int_{S_m} \sigma_m(y)dy. \qquad (11)$$

Using some estimates of $G(x,y)$ ([3]) and the asymptotic formula for $Q_m$ from [4], we can rewrite the exact formula (10) as follows:

$$u_M(x) = u_0(x) + \sum_{m=1}^{M} G(x,x_m)Q_m + o(1), \ a \to 0, \ |x - x_m| \geq a. \qquad (12)$$

The number $Q_m(x)$ is given by the asymptotic formula

$$Q_m = -4\pi h(x_m)u_e(x_m)a^{2-\kappa}[1+o(1)], \quad a \to 0, \qquad (13)$$

and the asymptotic formula for $\sigma_m$ is

$$\sigma_m = -\frac{h(x_m)u_e(x_m)}{a^{\kappa}}[1+o(1)], \quad a \to 0. \qquad (14)$$

Finally, formula for $u_e(x)$ is:

$$u_e^{(j)}(x) = u_0(x) - 4\pi \sum_{m=1, m \neq j}^{M} G(x, x_m)h(x_m)u_e(x_m)a^{2-\kappa}[1+o(1)]. \qquad (15)$$

Equation (9) for the limiting effective field $u(x)$ is used in numerical calculations when the number $M$ is very large, say $M = 10^b, b > 5$. The goal of our numerical experiments is to investigate the behavior of the solution to equation (9) and compare it with the asymptotic formula (15) in order to establish the limits of applicability of our asymptotic approach to many-body wave scattering problem for small particles.

## 4. Reduction of the scattering problem to solving linear algebraic systems

The numerical calculation of the field $u_e$ by formula (15) requires the knowledge of the numbers $u_m := u_e(x_m)$. These numbers are obtained by solving a linear algebraic system (LAS). This system is:

$$u_j = u_{0j} - 4\pi \sum_{m=1, m \neq j}^{M} G(x_j, x_m)h(x_m)u_m a^{2-\kappa}, \quad j = 1, 2, ..., M. \qquad (16)$$

This LAS is convenient for numerical calculations, because its matrix is often diagonally dominant. It follows from the results in [5], that for sufficiently small $a$ this system is uniquely solvable.

For finding the solution to the limiting equation (9), we use the collocation method from [5], which yields the following LAS:

$$u_j = u_{0j} - 4\pi \sum_{p=1, m \neq j}^{P} G(x_j, x_p)h(y_p)N(y_p)u_p |\Delta_p|, \quad p = 1, 2, ..., P, \qquad (17)$$

where $P$ is number of small cubes $\Delta_p$, $y_p$ is center of $\Delta_p$, $|\Delta_p|$ is volume of $\Delta_p$. We assume that the union of $\Delta_p$ forms a partition of $D$, and the diameter of $\Delta_p$ is $O(d^{1/2})$.

From the computational point of view solving LAS (17) is much easier than LAS (16), because $P << M$.

We have two different LAS, corresponding to formula (15) and to equation (9). Solving these LAS, one can compare their solutions and evaluate the limits of applicability of the asymptotic approach from [4] to solving many-body wave scattering problem in the case of small particles.

## 5. Numerical simulation

We take $k = 1$, $\kappa = 0.9$, and $N(x) = N = const = 10$ for the numerical calculations. For $k = 1$ and the values of $a$ and $d$, used in our numerical experiments, one can have many small particles on the wavelength. Therefore, the multiple scattering effects are not negligible.

The numerical procedure for checking the accuracy of the solution to equation (9) uses the calculations with various values of equation parameters. The absolute and relative errors were calculated by increasing the number of collocation points. The accuracy of the solution increases if the number of collocation points grows. The dependence of the accuracy on the parameter $\mathcal{P}$, where $\mathcal{P} = \sqrt[3]{P}$, $P$ is the total number of small subdomains in $D$, is shown in Fig. 1.

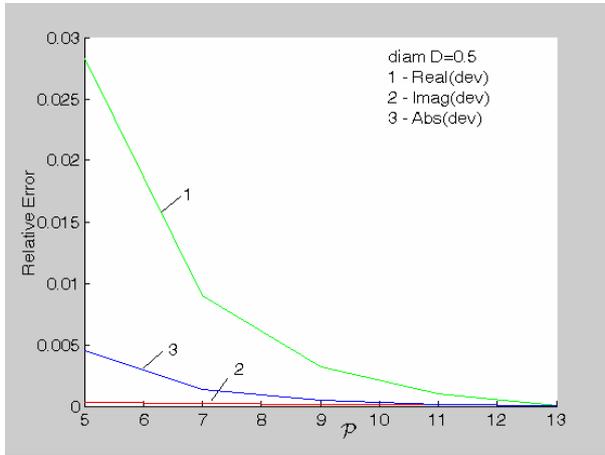 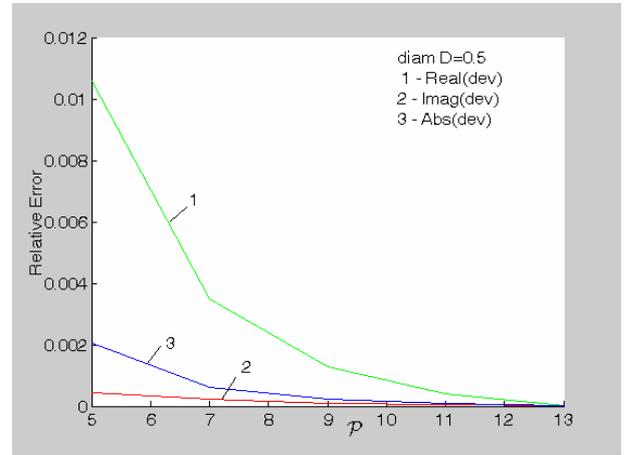

*Fig. 1a. Relative error versus the $\mathcal{P}$ parameter, $h(x) = k^2(1-7i)/(40\pi)$.*   *Fig. 1b. Relative error versus the $\mathcal{P}$ parameter, $h(x) = k^2(1-3i)/(40\pi)$.*

The error of the solution to equation (9) is equal to 1.1% and 0.02% for real and imaginary part respectively at $\mathcal{P}=5$ (125 collocation points), it decreases to values 0.7% and 0.01% if $\mathcal{P}=6$ (216 collocation points), and it decreases to values 0.29% and 0.005% if $\mathcal{P}=8$ (512 collocation points), $h(x) = k^2(1-3i)/(40\pi)$. The relative error less than 0.1% (at $\mathcal{P}=12$) is obtained in the process of numerical calculations. This error depends on the function $h(x)$ too.

The comparison of the solutions to LAS (16) and (17) was carried out for various values of $a$ and various values of the number $P$. The relative error of the solution decreases when $P$ grows. For example, when $P$ increases by 50%, the relative error decreases by 12% (for $\mathcal{P}=8$).

The difference between the real parts, imaginary parts, and moduli of the solutions to LAS (16) and (17) are shown in Fig. 2 for $\mathcal{P}=7$. The real part of this difference does not exceed 4% when $a = 0.01$, $N = 20$, it is less than 3.5% at $a = 0.008$, less than 2% at $a = 0.005$; $d = 8a$. This difference is less than 0.08% when $\mathcal{P}=11$, $a = 0.001$, and $d = 15a$.

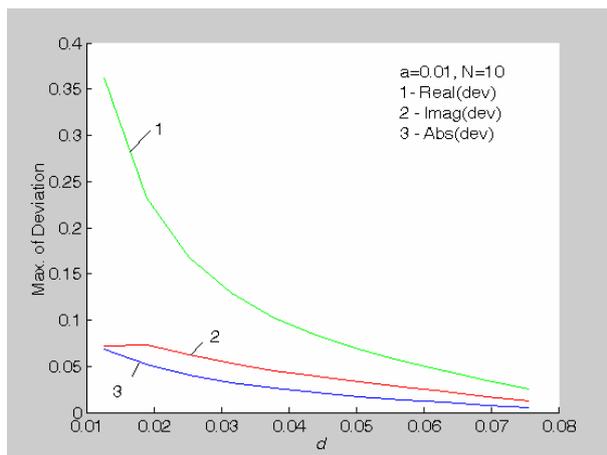 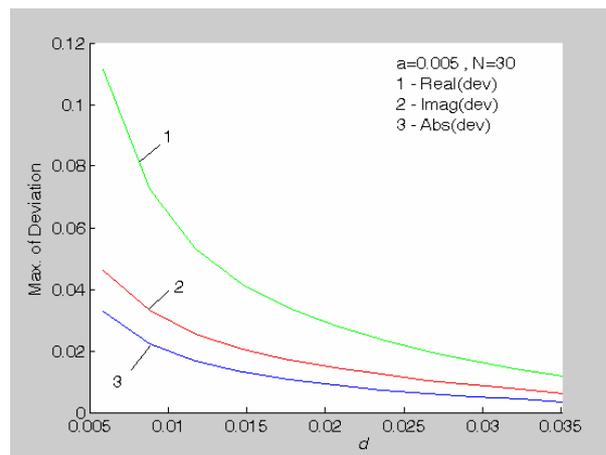

*Fig. 2a. Deviation of component field versus the distance $d$ between particles, $N = 10$.*

*Fig. 2b. Deviation of component field versus the distance $d$ between particles, $N = 30$.*